# Measurement of XUV-absorption spectra of ZnS radiatively heated foils.

N. Kontogiannopoulos[1], S. Bastiani-Ceccotti[1], F. Thais[2], C. Chenais-Popovics[1], P. Sauvan[3], R. Schott[1], W. Fölsner[4], Ph. Arnault[5], M. Poirier[2], T. Blenski[2]

[1]Laboratoire pour l'Utilisation des Lasers Intenses, UMR n°7605 CNRS–Ecole Polytechnique–CEA–Univ. Paris VI, 91128 Palaiseau Cedex, France
[2]DSM/DRECAM/SPAM, CEA-Saclay, 91191 Gif-sur-Yvette Cedex, France
[3] Departamento de Ingeniería Energética, UNED, Juan del Rosal 12, 28040 Madrid
[4] Max-Planck Institut für Quantenoptik, Garching, Germany
[5] Centre DAM-Ile-de-France, BP 12, 91680 Bruyères-le-Châtel, France

**Abstract**

Time-resolved absorption of zinc sulfide (ZnS) and aluminum in the XUV-range has been measured. Thin foils in conditions close to local thermodynamic equilibrium were heated by radiation from laser-irradiated gold spherical cavities. Analysis of the aluminum foil radiative hydrodynamic expansion, based on the detailed atomic calculations of its absorption spectra, showed that the cavity emitted flux that heated the absorption foils corresponds to a radiation temperature in the range 55 – 60 eV. Comparison of the ZnS absorption spectra with calculations based on a superconfiguration approach identified the presence of species $Zn^{6+}$ - $Zn^{8+}$ and $S^{5+}$ - $S^{6+}$. Based on the validation of the radiative source simulations, experimental spectra were then compared to calculations performed by post-processing the radiative hydrodynamic simulations of ZnS. Satisfying agreement is found when temperature gradients are accounted for.

1. **Introduction**

The energy exchange in stellar interiors is dominated by radiative transfer, which is determined by the absorption of medium-Z elements, even though they represent only a small fraction of the stellar mass [1,2]. This stems from the strong absorption structures of medium-Z elements in the X-ray and XUV range that match the Planckian spectrum of the radiated flux. Further, the study of absorption coefficients is of great interest for the indirect scheme of inertial confinement fusion, where the deposited energy is dominated by the radiative properties of the atomic species used in hohlraums and pellets layers [3].

In the present experiment the absorption coefficients of zinc sulfide (ZnS) and aluminum plasmas in local thermodynamic equilibrium (LTE) conditions were characterized by measurements of their absorption spectra. ZnS and Al foils were heated by the radiation emitted by a gold spherical cavity. The ZnS absorption spectra were analyzed by comparing the experimental measurements with the theoretical predictions of the SCO atomic physics code [4,5], based on the superconfiguration approximation [6].

Measurements of the absorption spectra of plasma mixtures are timely and important as they can provide benchmarks for recently developed theoretical models [7]. Such models are of particular interest especially in astrophysics where plasma mixtures dominate stellar behavior. The compound ZnS was chosen because the sulfur n=2 to 3 and zinc n=3 to 4 transitions, where n is the principal quantum number, lie in the same spectral region and do not overlap. Finally, the Al study was performed because its absorption spectra have been previously measured [8,2] and as a low-Z element its absorption spectra can also be accurately predicted by detailed atomic physics codes such as HULLAC [9,10]. Thus, it provides an independent means of inferring the foil temperature achieved under the specific experimental conditions.

## 2. Experimental setup and methods

2.1 Experimental setup

The experiment was performed at the LULI2000 laser facility, where two frequency-doubled ($\lambda$ = 0.53 µm), 500 ps pulse duration, Nd-glass laser beams were used. A schematic of the experimental setup is given in Fig. 1. The first beam, hereafter called the "main beam", heated the spherical gold cavity. It was focused with a $f$ = 800 mm lens coupled with a random phase plate (RPP) to obtain a 500 µm diameter FWHM focal spot at the entrance hole of the cavity. The energy of the main beam at 0.53 µm wavelength was ~130 J giving an average intensity of $1.3 \times 10^{14}$ W/cm$^2$.

A 1.2 mm diameter spherical gold cavity was used for the radiative heating of the absorption foils. The main beam heated a 130 nm thick gold foil located at the entrance hole of the cavity (Ø 700 µm). The radiation emitted from the back side of this foil was absorbed by the wall of the cavity producing a plasma layer on its interior, which then emitted radiation that was confined in the cavity. An absorption foil placed on one of the two diagnostic holes (Ø 250 µm) was heated by the radiation emitted from the cavity and by the directly-irradiated Au foil.

Two foils thicknesses were used: 48.9 nm or 97.8 nm for ZnS and 74 nm or 148 nm for Al, which correspond to 20 and 40 µg/cm$^2$ areal mass, respectively. The absorption foils of both types were coated on both sides with 35.3 nm thick carbon tampers corresponding to 8 µg/cm$^2$ areal mass. The tampered configuration was used to reduce the density and temperature gradients along the plasma expansion axis, and thus to improve the homogeneity of the probed plasma [11].

The second laser beam, delivering an energy of 30 J at 0.53 µm wavelength - hereafter called the "backlight beam", or BL beam - was focused on a 20 µm thick gold foil to produce a XUV absorption source, or backlight, used to probe the radiatively heated foils. The backlight producing gold foil was placed at a distance of 3 mm from the cavity to prevent the heating of the absorption foils. The BL beam was delayed by 800 ps with respect to the main beam. This delay ensures that the absorption probing occurs late enough to minimize the temperature gradient of the target plasma introduced by its hydrodynamic expansion, but early enough to prevent the plasma obscuring the line of sight along from the BL to the spectrometer defined by the cavity diagnostic holes (See Fig. 1). Furthermore, with this delay, the probing occurs when the radiative emission of the cavity is small, being on the order of the background level.

Both laser focal spots were monitored using two X-ray pinhole cameras. The backlight spectra were time-resolved using an XUV spectrograph coupled with a X-ray streak camera. The spectrograph was composed of a 2000 lines/mm gold transmission grating with a Ni spherical mirror with curvature radius 5200 mm. The mirror was placed with a grazing incidence angle of $6.4^o$ to reject the energetic X-ray photons with wavelengths below 20 Å. Due to the spectrograph dispersion, the streak camera recorded photons up to 200 Å. Thus, the spectrograph covered the spectral range 20 – 200 Å with a resolution of 2.4 Å. The time resolution provided by the streak camera was 50 ps.

2.2 Backlighter spectrum characterization

Figure 2 shows a raw time-resolved spectrum of a 20 µg/$cm^2$ Al absorption foil. The weak spectral feature emitted first corresponds to the self-emission of the cavity transmitted by the foil, while the second feature corresponds to the transmitted backlighter emission. The backlight spectra were obtained from the streak images with temporal average corresponding to an integration of 500 ps, in order to minimize the noise on the measured spectra.

As the spectra of the backlight and of the backlight absorbed by the foil were measured in different experiments, it was necessary to normalize the absorption spectra to the spectra of the backlight used in each shot and scaled from the laser energy. The gold emission is characterized by a fairly flat spectrum without discrete emission lines, except for a broad structure in the region around 50 Å (O – band transitions). The normalization was performed taking into account the linear dependence of the spectral intensity level as a function of the BL laser energy. The spectral range for the analysis was limited to 55-125 Å, as the O-band structure below 55 Å had a strong non-linear dependence on the BL laser energy, and the BL emission level was too low above 125 Å.

2.3 Method of analysis of the spectra

ZnS and Al spectra were analyzed by comparison of the experimental spectra with the results of the codes SCO [4,5,7] and HULLAC [9], respectively, used to post-process the output of the radiative hydrodynamic code MULTI [12]. The detailed atomic physics code HULLAC provides the transmission of individual ion stages. HULLAC uses the energy levels and the oscillator strengths for each single ion species calculated and tabulated by RELAC, which is based on the relativistic parametric potential method [10]. Then the plasma transmission is calculated using a Saha-Boltzmann distribution of the ion populations for the conditions provided by MULTI.

On the other hand, the version of SCO used here is based on the statistical theory for the description of the super-transition arrays in LTE plasmas [4-7]. SCO can calculate the absorption and transmission spectra of all elements separately as well as the total absorption and transmission spectrum of the mixture. The ion-in-cell model [4-6] is used for all elements in the mixture so that all the ions are screened by a plasma having the same electron density [7]. In the present work, the Thomas-Fermi model of mixtures is used self-consistently to find the effective mass density of each element corresponding to the free electron density. The ions of the mixture and of the carbon tamper are represented by hundreds of superconfigurations the structure of which are calculated self-consistently in a neutral atomic sphere corresponding to the effective mass density of the element [4,5]. The absorption spectra are calculated separately for each superconfiguration and the total absorption spectrum is the sum over all superconfiguration weighted by their relative probabilities [4,5,7]. For each superconfiguration SCO does not calculate detailed lines but uses instead a statistical approach both for terms [13] and for configurations [4-7] resulting in unresolved absorption structures (UTA or SOSA). For each one-electron transition between two configurations a continuous envelope with the UTA or SOSA width and center of gravity represents the array of corresponding detailed lines. These envelopes are averaged over all configurations belonging to the superconfiguration [4,5,6]. The orbital relaxation corrections have been included for all relevant transitions [5].

The determination of the hydrodynamic parameters, used by SCO and HULLAC, necessitates two simulations with the code MULTI because the radiative heat flux driving the absorption foil and the associated radiative temperature are not known. In a first simulation, the laser irradiated conversion foil expansion and the radiative flux emitted at the rear of this foil is calculated. Then the theoretical model proposed by Basko [14] is used to determine the energy balance in the cavity, which provides the radiative flux that heats the absorption foil. A second simulation is then performed to calculate the hydrodynamic behavior of the absorption foil. Aluminum spectra are used here to check the validity of the method and the evaluation of the radiative flux heating the absorption foil.

3. **Determination of the heating radiative flux from the analysis of the Al spectra**

3.1 Al absorption spectra

The analysis was performed for a 20 µg/cm$^2$ areal mass Al foil, where the main laser beam flux was 1.4x10$^{14}$ W/cm$^2$ at 0.53 µm. The experimental absorption spectrum averaged over the probing pulse duration is given in Fig. 3 where the dominant transitions are due to the 2$p$ – 3$d$ transitions of Al$^{5+}$ to Al$^{7+}$, at 68, 76 and 88 Å, respectively. The Al$^{5+}$ 2$p$ – 3$s$ transitions are observed at 105 and 110 Å. The NIST tables [15] suggest that the transition close to 96 Å may originate in Al$^{6+}$ 2$p$ – 3$s$ transitions, but the HULLAC transitions strengths are weak for these lines, so additional spectroscopic data would be useful. This ionization distribution, broader than those in previous works [8], suggests the presence of gradients - most likely in the temperature - in the heated Al.

3.2 Calculation of the radiation flux heating the aluminum foil

The Basko model, used to calculate the cavity heating and flux on the Al foil, is based on the modified asymptotic self-similar solution for the heating of a planar medium, in conjunction with a view factor approach for the description of the energy balance inside the cavity [14, 16-18].

The X-ray flux emitted at the rear of the conversion foil was calculated in a first simulation with MULTI. A 130 nm thick gold foil was heated on one side (front) by a Gaussian laser pulse of 500 ps duration and 1.4x10$^{14}$ W/cm$^2$ flux at 0.53 µm wavelength. This radiative flux was then used as an input to the Basko model, where the reemission factor (albedo) of the cavity wall was found equal to 0.3. Taking into account the view factors between the absorption foil, the conversion foil and the cavity wall, the flux emitted by the cavity wall heating the absorption foil was estimated to be 2% of the total flux emitted by the conversion foil, while 4% of the total flux emitted by the conversion foil directly heats the absorption foil, i.e., twice the flux emitted by the cavity which is comparable to the results reported in Ref. [19]. The total heating flux, characterized by the sum of the two fluxes, corresponds to 6% of the total flux emitted by the conversion foil. The equivalent temperature of the total incident flux on the absorption foil corresponds to a radiative temperature of 60 eV. Note that the spectrum emitted by the conversion foil deviates from a Planckian distribution so that an equivalent temperature is a valid concept for the total flux only and not for its spectral distribution [11,20].

3.3 Hydrodynamic simulation of the aluminum foil expansion

To simulate with MULTI the hydrodynamic expansion of the Al foil, the X-ray emission from the rear of the gold foil calculated above was used as the heating source. This choice was justified because the contribution of the conversion foil flux is dominant for the heating of the absorption foil. Furthermore, it permits one to account for the spectral distribution of the heating radiation to give a more accurate simulation than that using a Planckian source [11,20]. To this end, 6 % of the X-ray flux emitted at the back side of a 130 nm thick gold foil, obtained by MULTI for a 500 ps duration Gaussian pulse with $1.4 \times 10^{14}$ W/cm$^2$ flux at 0.53 μm wavelength, was used to irradiate a 20 μg/cm$^2$ Al foil, coated on each side with 8 μg/cm$^2$ carbon tampers. In Fig. 4 are the electron temperature and the matter density as a function of the Lagrangian mass at three times corresponding to the rising half-maximum (1.2 ns), maximum (1.4 ns) and falling half-maximum (1.7 ns) of the probing pulse. Those profiles cover the measurement integration interval of 500 ps, and permit one to follow the temporal evolution and spatial distribution of the hydrodynamics. We observe in Fig. 4(a) that the temperature spatial gradients of Al are initially large (22 – 30 eV at 1.2 ns and 20 – 28 eV at 1.4 ns) and smooth out as time evolves (18 – 23 eV at 1.7 ns). In Fig. 4(b), we see that the matter density spatial variations are 3 – 6 mg/cm$^3$ at 1.2 ns, 2.5 – 4 mg/cm$^3$ at 1.4 ns, and 2 – 3 mg/cm$^3$ at 1.7 ns, respectively.

3.4 Calculation of aluminum absorption spectra – Comparison with the measured absorption.

The hydrodynamic parameters of the Al foil given by MULTI were used to calculate *ab initio* the Al absorption spectra with the detailed code HULLAC. The spatial variation of the Al electron temperature was represented by dividing each time profile into a number of constant temperature cells using discrete steps of 2 eV. Because calculations with HULLAC showed that the matter density does not affect significantly the structures of the absorption of Al, a mean value of 2 mg/cm$^3$ was used as a first approximation. The carbon foils introduce only a lowering of the transmission due to the continuum absorption. Their contributions were taken into account to provide an absolute value of the absorption. Thus, for the tampers the spatial variation of their electron temperature was discretized using 4 eV steps, while their matter density was fixed to a mean value of 1 mg/cm$^3$. Then, the total transmission for each time profile was calculated by the product:

$$\tau(t_n) = \prod_i e^{-K_i \sigma_i}$$

where $K_i(\rho, T_i)$ is the opacity of the *i*-th spatial cell given by HULLAC for the mean density $\rho$ and the cell constant temperature $T_i$; $\sigma_i$ is the *i*-th cell areal mass, and the product includes both Al and carbon cells. The temporal variation, for the interval covering the time integration of the measured absorption, was approximated by calculating the average of the three time profiles using the formula

$$\tilde{\tau} = \frac{1}{3}\sum_{n=1}^{3}\tau(t_n)$$

where the time profiles are considered at $t$ = 1.2 ns, 1.4 ns and 1.7 ns, respectively.

The averaged transmission is shown in Fig. 5 together with the measured transmission. The theoretical transmission reproduces with good accuracy the measured structures corresponding to the $Al^{5+}$ and $Al^{6+}$ transitions identified in Fig. 5. However, below 75 Å the theoretical absorption level is systematically lower than the measured one. This is particularly obvious for the structure around 68 Å corresponding to the transitions of $Al^{7+}$ and from 100 to 110 Å where $Al^{4+}$ ions contribute to the absorption. These discrepancies may be attributed to the deviations of the ion populations from the Saha-Bolztmann statistics that have been reported in similar works [19,20].

This analysis of the Al spectra suggests that the equivalent temperature of 60 eV given by the Basko model can be an adequate estimate of the radiative flux heating the Al absorption foil. Taking into account the energy variations of the main beam, according to the Basko model, the equivalent temperatures of the heating flux are expected to be in the range of 55 – 60 eV for the different experimental shots.

4. **ZnS absorption spectra analysis**

The ZnS transmission spectra were analyzed with SCO. First, SCO permitted the identification of the atomic transitions of the various zinc and sulfur observed ion species. Second, SCO was coupled as a post-processor of MULTI output to calculate the transmission spectra of the ZnS plasma taking into account its spatial distribution and its time evolution.

In Fig. 6 is presented the measured transmission spectrum of a 20 µg/cm² ZnS foil. The main beam flux on the conversion foil was $10^{14}$ W/cm² at 0.53 µm. In the spectral range between 55 – 120 Å we observed various absorption structures corresponding to the transitions of both zinc and sulfur ions. In particular, the $3d – 4f$ transitions of $Zn^{6+}$ and $Zn^{7+}$ are dominant in the range 103 – 120 Å. The $3d – 4f$ transitions of $Zn^{8+}$ appear as two structures between 84 – 90 Å and 90 – 103 Å. At lower wavelengths $2p – 3s$ transitions of the ions $S^{5+}$ - $S^{6+}$ have been identified as the narrow structure observed between 70 – 75 Å.

For the experimental conditions of Fig. 6, according to the Basko model, the radiative flux heating the ZnS foil corresponds to an equivalent planckian radiation of 55 eV. Following the same procedure as for Al, the hydrodynamic simulation of ZnS was performed coupling directly the radiation emitted at the back side of a gold foil, rather than using a planckian driver.

The electron temperature time evolution of ZnS is given in Fig 7. During the probing pulse, the temperature variation of each cell (ZnS center cell and C/ZnS interfaces) is about 5 eV. The electron temperature and the matter density as a function of the Lagrangian mass at three times corresponding to the rising half-maximum (1.2 ns), the maximum (1.4 ns) and the falling half-maximum (1.7 ns) of the probing pulse are given in Fig. 8. According to Figure 8(a), the temperature spatial gradients are initially 17 – 20 eV at 1.2 ns and smooth out as time evolves to 14 – 16 eV at 1.4 ns and 13 – 14 eV at 1.7 ns. That is, MULTI predicts that the temperature gradients are mainly due to the time evolution. Further, in Fig. 8(b), the matter density presents a spatial variation of 4 – 5 mg/cm³ at 1.2 ns, of 3 – 4 mg/cm³ at 1.4 ns, and of 2.5 – 3 mg/cm³ at 1.7 ns, respectively.

The results calculated by MULTI have been post-processed with SCO to determine the transmission as was done for Al case. As a first approximation, the calculations were performed using a constant matter density of 2.5 mg/cm$^3$ for ZnS and of 1 mg/cm$^3$ for carbon, respectively.

The theoretical results are compared to the measured transmission in Fig. 9, where good agreement between the two is found in the range 84 – 120 Å, corresponding to the $3d – 4f$ transitions of $Zn^{6+} – Zn^{8+}$. However, there is a difference in the range 76 – 84 Å. In particular, the measured transmission has an absorption feature around 80 Å that does not appear in the theoretical transmission. Identification of supertransitions arrays with SCO show that this feature could correspond to the $Zn^{8+}$ $3d – 5f$ transitions. For these conditions the theory seems to follow only the weak narrow structure in the range 84 – 90 Å, which correspond to the $3d – 4f$ transitions of $Zn^{8+}$. For the absorption of the sulfur ions $S^{5+}$ - $S^{6+}$, the theoretical transmission is in agreement with the narrow absorption feature of the $2p – 3s$ transitions between 70 – 75 Å; however, a large deviation is observed in the absorption feature of the $2p – 3d$ transitions in the range 58 – 70 Å.

We note that the ZnS matter density used for the calculations is a first approximation. Calculations with SCO showed that the zinc transmission spectrum is sensitive to the density. This dependence affects the absorption and could explain the observed deviations of the absolute value of the transmission. Further, and more interestingly, the matter density affects the shape of the transmission spectra, in particular as concerns the $Zn^{8+}$ $3d – 5f$ transitions. This needs further analysis. The deviation for sulfur around 60 Å is probably due to the lower experimental accuracy at low wavelengths. Further analysis using HULLAC is needed to clarify this situation for the transmission [21].

## 5. Conclusions

The time resolved XUV absorption spectra of ZnS and Al foils heated by the thermal radiation of a gold cavity have been measured. Analysis of the Al absorption spectra with HULLAC post-processing the hydrodynamic output of MULTI, showed that the Planckian equivalent temperatures of the radiative flux heating the absorption foil was in the range of 55 – 60 eV. Based on this estimate of the cavity heating conditions, the ZnS foil hydrodynamic expansion was simulated with MULTI. The results were used by SCO to calculate the transmission spectra of ZnS. Comparison showed the dominance of the $3d – 4f$ transitions of $Zn^{6+}$ and $Zn^{7+}$ ions. Also, spectral features corresponding to the $3d – 4f$ and $3d – 5f$ transitions of $Zn^{8+}$ were identified. The deviations of the theoretical transmission in the range corresponding to $Zn^{8+}$ $3d – 5f$ transitions suggests a correlation between the matter density and the occurrence of the transitions, an interesting point which needs further exploration. For sulfur, the $2p – 3s$ transitions of $S^{5+}$ - $S^{6+}$ ions have been identified. The observed deviation between experiment and theory in the range corresponding to the $2p – 3d$ transitions will be further analyzed with the detailed code HULLAC.

**Acknowledgements**

The authors acknowledge the invaluable support of the LULI laser operations staff, in particular Marc Rabec-Le-Gloahec, Luc Martin and A. M. Sautivet. A special thank to the Max Planck Institute members, Klaus Eidmann and George Tsakiris for helpful discussions and specially to Walter Foelsner who was in charge of the hohlraum targets. T. Blenski, M. Poirier and F. Thais acknowledge a partial support from EURATOM.

# Figures Captions

**Figure 1 :** Experimental setup for the absorption measurement of ZnS foils heated by the X-rays radiative emission of a gold cavity.

**Figure 2 :** Time-resolved Al XUV-absorption image recorded on the streak camera.

**Figure 3 :** Measured transmission spectra of a 20 µg/cm$^2$ aluminum foil, tamped with 8 µg/cm$^2$ carbon foils on both sides. Identification of the plasma ionic species and atomic transitions is given.

**Figure 4 :** Multi simulation results of a 20 µg/cm$^2$ aluminum foil tamped with 8 µg/cm$^2$ carbon foils on both sides. (a) Electron temperature and (b) density as function of lagrangian mass at the rising half-maximum (1.2 ns), at the maximum (1.4 ns) and at the falling half-maximum (1.7 ns) of the probing pulse.

**Figure 5 :** Comparison of the experimental transmission (thin dashed line) with the averaged transmission calculated with HULLAC code (thick full line). The density and the electron temperature are calculated with MULTI. The time evolution gradients are included calculating separately the transmissions given by the spatial gradients distributions at three times and taking their average.

**Figure 6 :** Measured transmission spectra of a 20 µg/cm$^2$ ZnS foil, tamped with 8 µg/cm$^2$ carbon foils on both sides. Identification of the plasma ionic species and of the corresponding atomic transitions structures.

**Figure 7 :** MULTI simulation results of the temperature time evolution of 20 µg/cm$^2$ ZnS foil tamped with 8 µg/cm$^2$ carbon foils on both sides, at three interfaces covering the plasma spatial expansion. Dashed line : right carbon/ZnS interface. Full line : Interface at the center of ZnS. Dotted line : left carbon/ZnS interface.

**Figure 8 :** Lagrangian representation of plasma hydrodynamic parameters. (a) Electron temperature and (b) density at the rising half-maximum (1.18 ns), at the maximum (1.43 ns) and at the falling half-maximum (1.72 ns) of the probing pulse.

**Figure 9 :** Comparison of the experimental transmission (thin dashed line) with the averaged transmission calculated with the SCO code (thick full line). The density and the electron temperature are calculated with MULTI. The time evolution gradients are included calculating separately the transmissions given by the spatial gradients distributions at three times and taking their average.

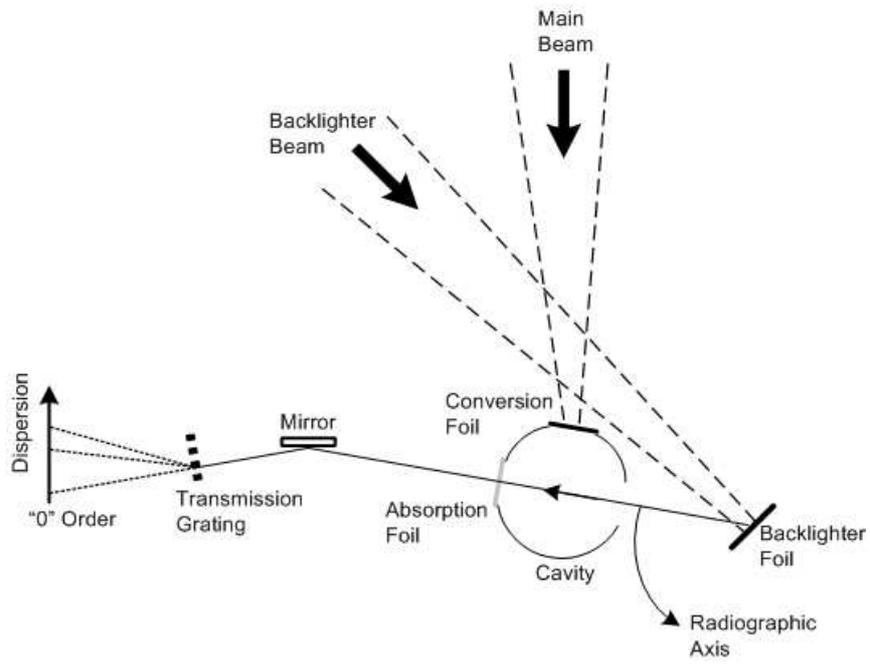

N. Kontogiannopoulos et al, Figure 1

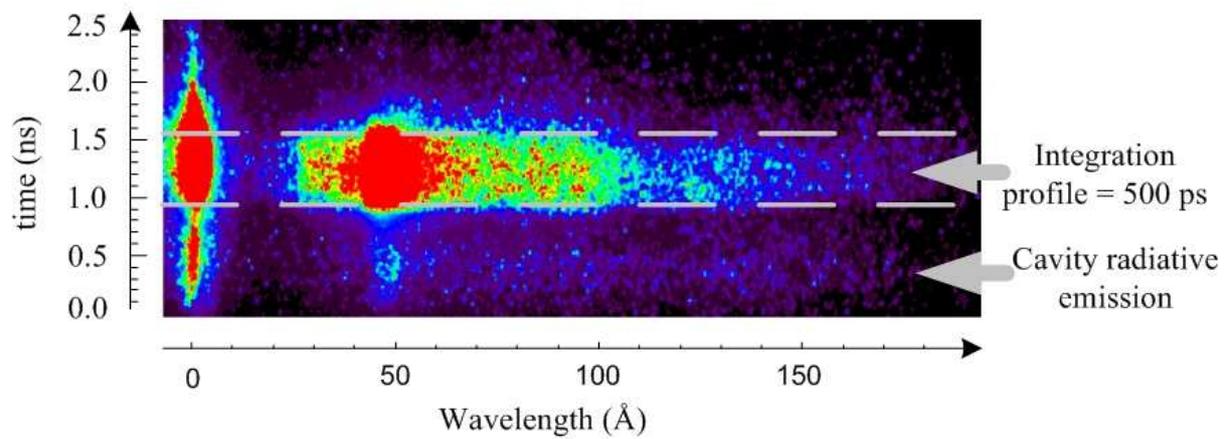

N. Kontogiannopoulos et al, Figure 2

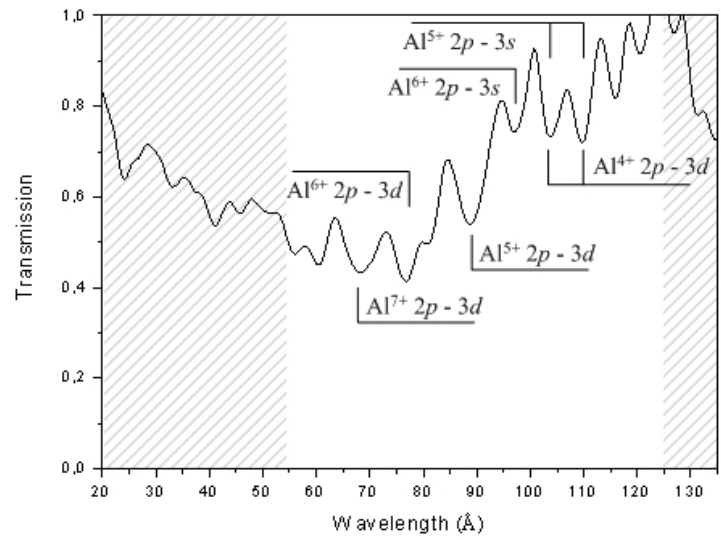

N. Kontogiannopoulos et al, Figure 3

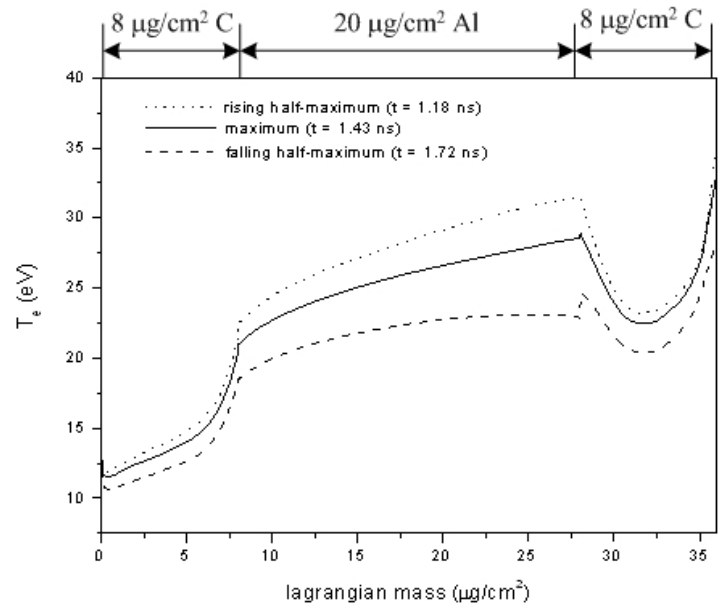

(a)

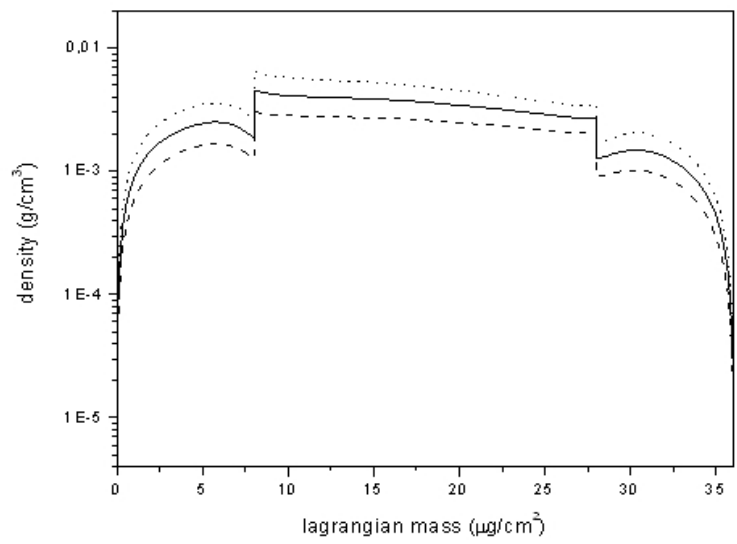

(b)

N. Kontogiannopoulos et al, Figure 4

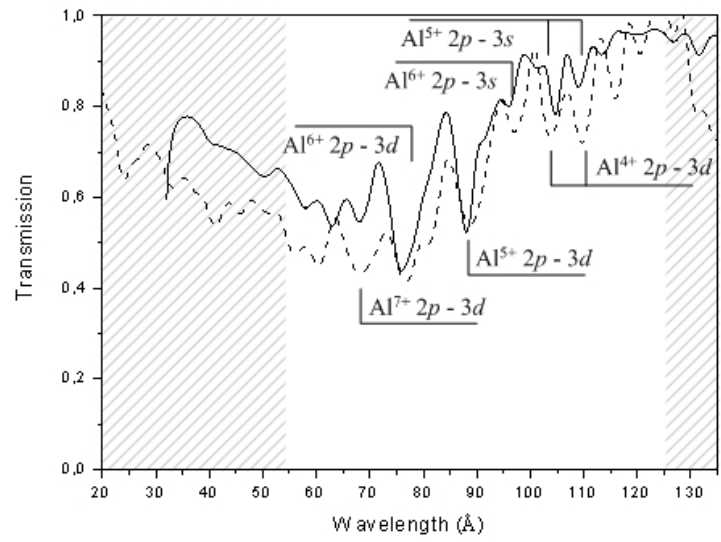

N. Kontogiannopoulos et al, Figure 5

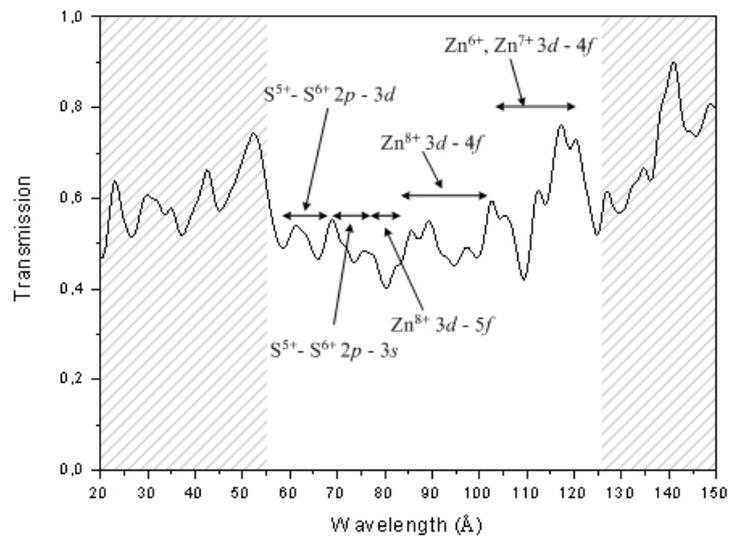

N. Kontogiannopoulos et al, Figure 6

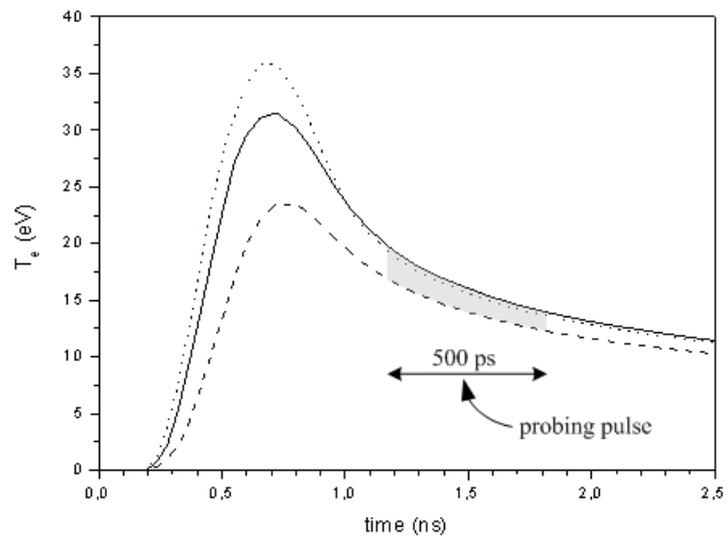

N. Kontogiannopoulos et al, Figure 7

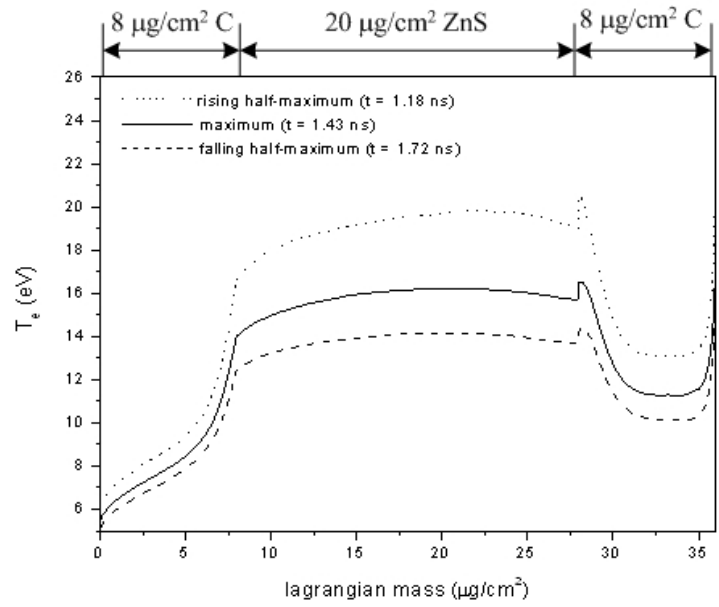

(a)

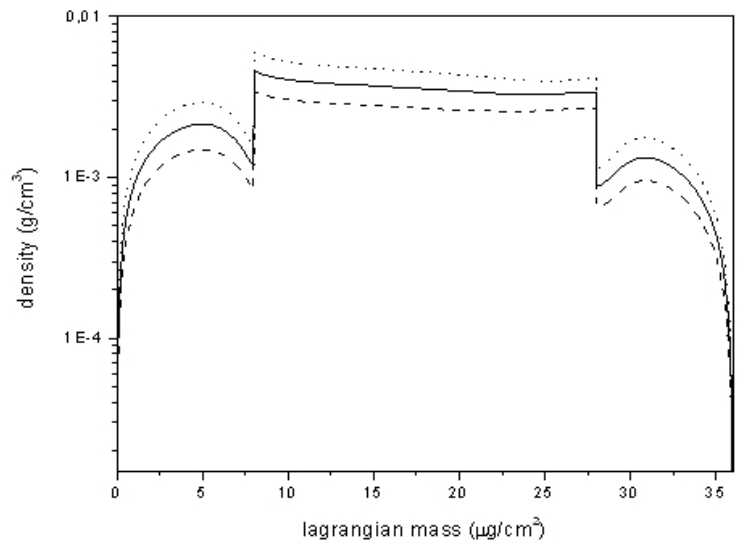

(b)

N. Kontogiannopoulos et al, Figure 8

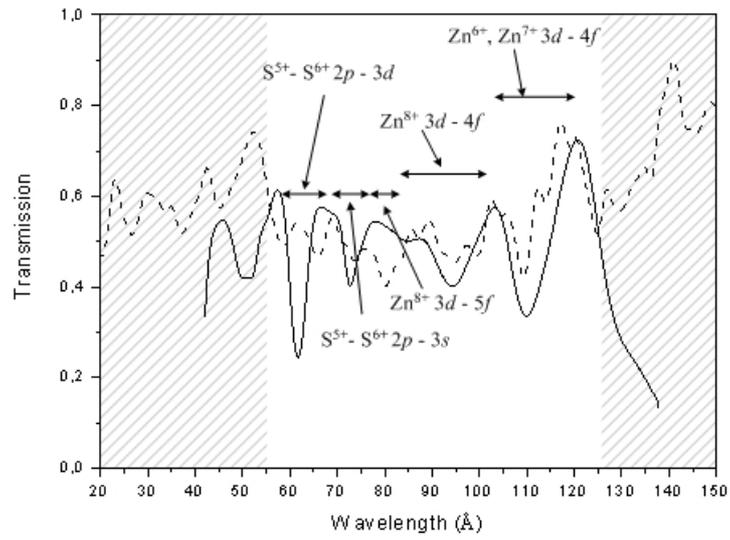

N. Kontogiannopoulos et al, Figure 9